\documentclass[conference]{IEEEtran}
\IEEEoverridecommandlockouts
\usepackage{cite}
\usepackage{amsmath,amssymb,amsfonts}
\usepackage{algorithmic}
\usepackage{graphicx}
\usepackage{textcomp}
\usepackage{xcolor}
\def\BibTeX{{\rm B\kern-.05em{\sc i\kern-.025em b}\kern-.08em
    T\kern-.1667em\lower.7ex\hbox{E}\kern-.125emX}}
\begin{document}

\title{Identifying Alzheimer’s Disease Prediction Strategies of Convolutional Neural Network Classifiers using R2* Maps and Spectral Clustering\\
\thanks{This study was funded by the Austrian Science Fund (FWF grant numbers: P30134, P35887).}
}

\author{\IEEEauthorblockN{Christian Tinauer}
\IEEEauthorblockA{\textit{Department of Neurology} \\
\textit{Medical University of Graz}\\
Graz, Austria \\
christian.tinauer@medunigraz.at}
\and
\IEEEauthorblockN{Maximilian Sackl}
\IEEEauthorblockA{\textit{Department of Neurology} \\
\textit{Medical University of Graz}\\
Graz, Austria \\
maximilian.sackl@medunigraz.at}
\and
\IEEEauthorblockN{Stefan Ropele}
\IEEEauthorblockA{\textit{Department of Neurology} \\
\textit{Medical University of Graz}\\
Graz, Austria \\
stefan.ropele@medunigraz.at}
\and
\IEEEauthorblockN{Christian Langkammer}
\IEEEauthorblockA{\textit{Department of Neurology} \\
\textit{Medical University of Graz}\\
Graz, Austria \\
christian.langkammer@medunigraz.at}
}

\maketitle

\begin{abstract}
Deep learning models have shown strong performance in classifying Alzheimer's disease (AD) from R2* maps, but their decision-making remains opaque, raising concerns about interpretability. Previous studies suggest biases in model decisions, necessitating further analysis. This study uses Layer-wise Relevance Propagation (LRP) and spectral clustering to explore classifier decision strategies across preprocessing and training configurations using R2* maps. We trained a 3D convolutional neural network on R2* maps, generating relevance heatmaps via LRP and applied spectral clustering to identify dominant patterns. t-Stochastic Neighbor Embedding (t-SNE) visualization was used to assess clustering structure. Spectral clustering revealed distinct decision patterns, with the relevance-guided model showing the clearest separation between AD and normal control (NC) cases. The t-SNE visualization confirmed that this model aligned heatmap groupings with the underlying subject groups. Our findings highlight the significant impact of preprocessing and training choices on deep learning models trained on R2* maps, even with similar performance metrics. Spectral clustering offers a structured method to identify classification strategy differences, emphasizing the importance of explainability in medical AI.
\end{abstract}


\section{Introduction}
Deep neural networks have demonstrated strong performance in Alzheimer’s disease (AD) classification using magnetic resonance imaging (MRI) \cite{wen_convolutional_2020}, but their decision-making processes remain largely opaque \cite{davatzikos_machine_2019}. Ensuring that spurious data artifacts do not drive model accuracy is crucial for medical applications. While explainability methods such as Integrated Gradients \cite{sundararajan_axiomatic_2017}, LIME \cite{ribeiro_why_2016}, and Layer-wise Relevance Propagation (LRP) \cite{bach_pixel-wise_2015} have been used to highlight relevant regions in MRI-based classification, it remains unclear whether these networks primarily rely on disease-related biomarkers or unintended image characteristics \cite{tinauer_interpretable_2022}.

In this work, we utilized the effective relaxation rate R2* as a quantitative MRI parameter for AD classification with a relevance-regularized convolutional neural network (CNN). R2* is defined as the inverse of the effective transverse relaxation time T2* (R2*=1/T2*), which reflects the decay of transverse magnetization in each voxel. Notably, R2* is highly correlated with iron concentration in gray matter \cite{damulina_cross-sectional_2020}, and increased iron levels in the basal ganglia are frequently observed in AD \cite{drayer_imaging_1988}. We hypothesize that CNNs implicitly learn such patterns and, with recent advances in explainability, we can now disentangle and visualize these learned features.

By applying Spectral Relevance Analysis (SpRAy) \cite{lapuschkin_unmasking_2019} to LRP-based heatmaps, we systematically investigate the spatial clustering of relevance in a deep learning model trained on R2* maps. This approach allows us to identify dominant feature clusters that drive classification decisions and assess their consistency across preprocessing variations. Our findings provide deeper insights into how CNNs utilize structural brain information, further refining the understanding of preprocessing influences and potential biases in deep learning-driven AD classification.

\section{Methods}

\subsection{Dataset}

We retrospectively selected 226 MRI datasets from 117 patients with probable 
AD (mean age=71.1±8.2 years, male/female=93/133) from our outpatient
clinic and 226 MRIs from 219 propensity-logit-matched (covariates: age, sex)
\cite{kline_psmpy_2022,rosenbaum_central_1983} normal controls (NCs) (mean age=69.6±9.3 years, m/f=101/125) from an ongoing community-dwelling study. MRI data were acquired longitudinally over multiple sessions using a consistent MRI protocol at 3 Tesla (Siemens TimTrio), including a structural T1-weighted MPRAGE sequence with 1mm isotropic resolution (TR/TE/TI/FA =
1900 ms/2.19 ms/900 ms/9°, matrix = $176 \times  224 \times 256$) and a spoiled FLASH
sequence ($0.9 \times 0.9 \times 2$mm³, TR/TE=35/4.92ms, 6 echoes, 4.92ms echo spacing,
matrix = $208 \times 256 \times 64 \times 6$).

\subsection{Preprocessing}

Brain masks for each subject were obtained using FSL-SIENAX \cite{smith_accurate_2002}, and the
structural T1-weighted image, and were subsequently used to perform skull-stripping. Using the data acquired from the spoiled FLASH
sequence, we solved the inverse problem given as
\begin{equation}
    S_{xy}[i] = S_{xy}[0]e^{-t[i]R_2^*},
\end{equation}

for $S_{xy}[0]$ and $R_2^*$, where $S_{xy}[i]$ the measured voxel signal intensity at echo $i$ with echo time $t[i]$. The computations were executed for
all voxels in the image volumes, yielding the R2* maps (matrix = $208 \times 256 \times 64$).
The obtained R2* maps were affinely registered to the subject's MPRAGE
sequence using FSL-flirt and subsequently nonlinearly registered to the MNI152 standard-space 
brain template using FSL-fnirt.

\subsection{Classifier Network and Training}
We employed a 3D subject-level classifier network based on \cite{wen_convolutional_2020}, reducing the number and size of convolutional and fully connected layers to mitigate overfitting while maintaining validation accuracy. Batch normalization had no impact and was omitted, while max pooling was replaced with strided convolutions for improved interpretability \cite{springenberg_striving_2015, montavon_explaining_2017}. To enhance sparsity, all biases were constrained to be negative \cite{montavon_explaining_2017}.

The final architecture consists of four blocks, each containing a $3 \times 3 \times 3$ convolutional layer ($8$ channels) and a down-convolutional layer (strided $2$). This is followed by two fully connected layers ($16$ and $2$ units, respectively), totaling 0.3 million trainable parameters. ReLU activations were used throughout, except for the Softmax output layer.

To focus the network on relevant features, we implemented a relevance-guided architecture, $Graz^+$, which extends the classifier with a relevance map generator. This approach incorporates an additional loss term that encourages the model to assign higher relevance to predefined focus regions while suppressing irrelevant areas. The binary attention masks used for this guidance were derived from the FSL-SIENAX brain masks during preprocessing. Full methodological details can be found in \cite{tinauer_interpretable_2022}.

We trained models on R2* maps in subject space using the Adam optimizer \cite{kingma_adam_2017} for 60 epochs with a batch size of 6. Data was split into training, validation, and test sets (70:15:15) while ensuring that all scans from the same subject remained in the same set. To maintain class balance, final sets were constructed by combining subsets from each cohort, and the process was repeated $10$ times for random sampling analysis \cite{bradshaw_guide_2023}. The learning rate was initially set to $10^{-3}$, reduced by $0.3$ after five consecutive epochs without validation loss improvement, and had a lower bound of $10^{-6}$. Model weights were reset to their state at the start of a plateau phase.

\subsection{Model Configurations}

We evaluated three model configurations to assess the impact of skull-stripping and relevance-guided training \cite{tinauer_interpretable_2022} on classification performance. This design enabled us to isolate the effects of each component and analyze the learned features using heatmapping.

\subsection{Heatmapping}
Heatmaps were created using the LRP method with $\alpha=1.0$ and $\beta=0.0$, as described in \cite{bach_pixel-wise_2015}. Each voxel is attributed a relevance score (R). To analyze the relevance heatmaps, we grouped them and calculated mean heatmaps for each group.

\subsection{Spectral Relevance Analysis}

Spectral relevance analysis (SpRAy) enables efficient exploration of classifier behavior across large datasets by applying spectral clustering to inputs and heatmaps. This technique identifies common and atypical decision-making patterns, highlighting image features that may or may not reflect clinically relevant concepts. SpRAy is helpful in uncovering unexpected or artifact-driven classifier behaviors, similar to the "Clever Hans" effects found in \cite{lapuschkin_unmasking_2019,tinauer_pfungst_2025}.

The SpRAy process implemented for this study involves six steps:
\begin{enumerate}
\item Compute relevance maps using LRP to identify focus areas for classification.
\item Warp the heatmaps to MNI152 image space.
\item Downsample the native and warped heatmaps to 2 mm isotropic resolution for efficient analysis.
\item Perform spectral clustering to group similar image- or relevance patterns.
\item Identify clusters with highest eigenvalue gap, indicating well-separated heatmap groups and computing mean heatmaps for groups.
\item Visualize the clusters using t-Stochastic Neighbor Embedding (t-SNE) \cite{maaten_visualizing_2008}, which aids in interpreting the results and understanding the relationship between clusters.
\end{enumerate}

\section{Results}

Table~\ref{tab1} summarizes the performance metrics (accuracy, sensitivity, specificity, and AUC) for all configurations in the random sampling setup, evaluated in nonexcluded training sessions. Model A uses native R2* maps, Model B applies the brain mask to R2* maps for skull-stripping before classification, and Model C combines native R2* maps and relevance-guided training with brain masks.

\begin{table*}[htbp]
\caption{
Performance metrics for all model configurations on the AD vs NC classification task. Models are identified by Id, with columns indicating use of skull-stripping and relevance-guided training. Brackets [ ] show 95\% confidence intervals. NC: normal control; AD: Alzheimer's disease; AUC: area under the receiver operating characteristic curve
}
\begin{center}
\begin{tabular}{|l|l|l|l|l|l|l|}
\hline
\textbf{Id} & \textbf{Skull-stripping} & \textbf{Relevance-guided} & \textbf{Accuracy}                                                         & \textbf{Sensitivity}                                                       & \textbf{Specificity}                                                      & \textbf{AUC}                                                          \\ \hline
A           & No                       & No                        & \begin{tabular}[c]{@{}l@{}}64.2±6.5\%\\ {[}53.5\%, 76.9\%{]}\end{tabular} & \begin{tabular}[c]{@{}l@{}}61.3±9.6\% \\ {[}39.5\%, 74.7\%{]}\end{tabular} & \begin{tabular}[c]{@{}l@{}}67.0±9.4\%\\ {[}51.2\%, 83.7\%{]}\end{tabular} & \begin{tabular}[c]{@{}l@{}}64.12±0.06\\ {[}0.53, 0.77{]}\end{tabular} \\ \hline
B           & Yes                      & No                        & \begin{tabular}[c]{@{}l@{}}77.0±5.8\%\\ {[}64.9\%, 85.7\%{]}\end{tabular} & \begin{tabular}[c]{@{}l@{}}75.1±7.8\%\\ {[}62.2\%, 86.4\%{]}\end{tabular}  & \begin{tabular}[c]{@{}l@{}}78.8±7.9\%\\ {[}62.9\%, 90.1\%{]}\end{tabular} & \begin{tabular}[c]{@{}l@{}}76.94±0.06\\ {[}0.65, 0.85{]}\end{tabular} \\ \hline
C           & No                       & Yes                       & \begin{tabular}[c]{@{}l@{}}75.9±5.1\%\\ {[}67.9\%, 85.8\%{]}\end{tabular} & \begin{tabular}[c]{@{}l@{}}69.7±9.7\%\\ {[}52.9\%, 86.5\%{]}\end{tabular}  & \begin{tabular}[c]{@{}l@{}}81.6±5.5\%\\ {[}72.2\%, 92.3\%{]}\end{tabular} & \begin{tabular}[c]{@{}l@{}}75.64±0.05\\ {[}0.68, 0.86{]}\end{tabular} \\ \hline
\end{tabular}
\label{tab1}
\end{center}
\end{table*}

\begin{figure*}[htbp]
\centerline{\includegraphics[width=0.98\textwidth]{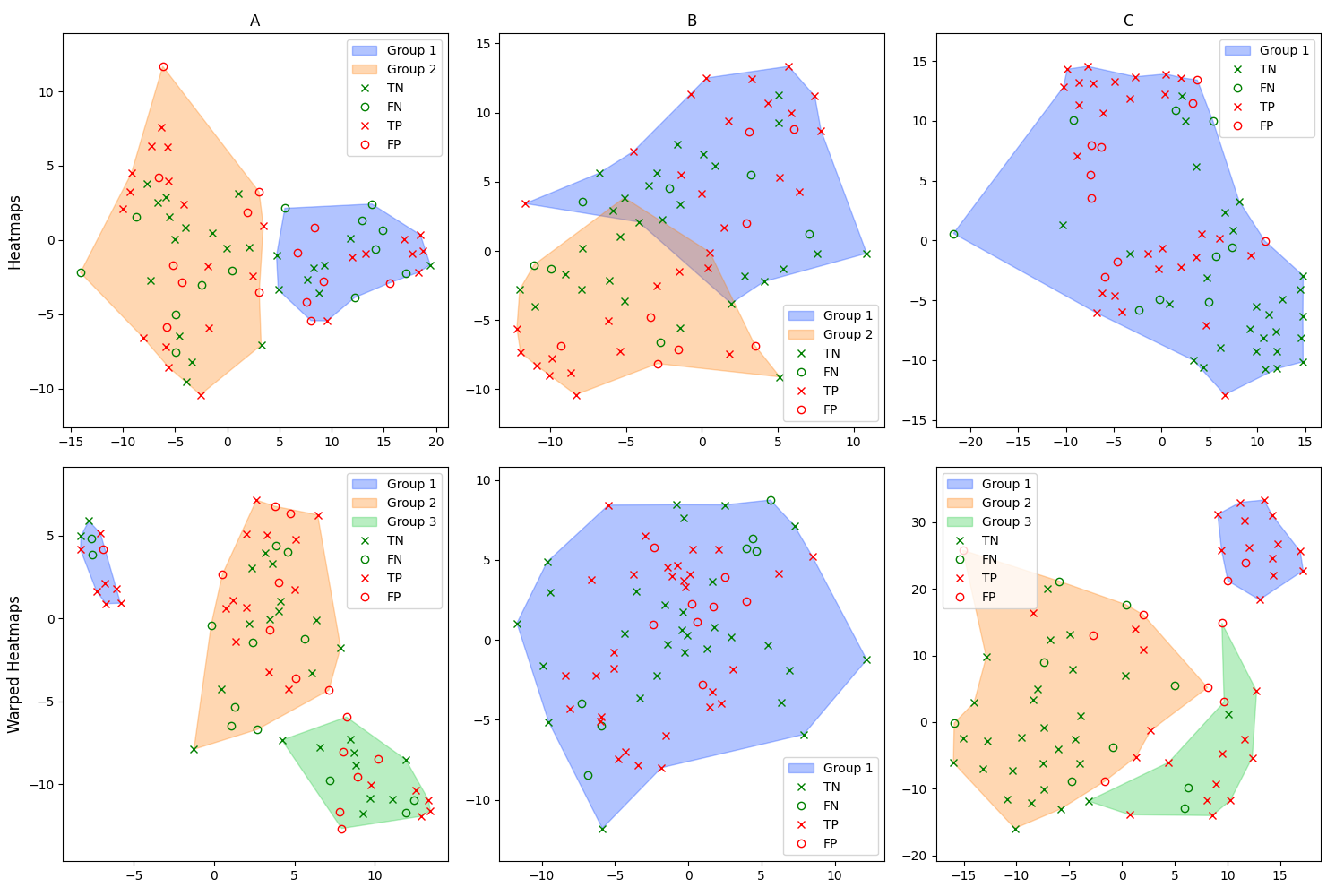}}
\caption{t-SNE visualization of relevance heatmaps (row 1) and warped heatmaps (row 2) for models trained on native R2* maps (A), skull-stripped R2* maps (B), and with relevance-guided training (C). Spectral clustering was used to group the heatmaps, with the number of clusters determined by eigenvalue analysis. Points indicate individual heatmaps and are colored by classification outcome (TN, FN, TP, FP). Only the warped heatmaps from model C show clustering that aligns with subject groups (NC vs. AD), indicating spatially distinct feature patterns. NC = normal control; AD = Alzheimer’s disease; TN = true negative; FN = false negative; TP = true positive; FP = false positive}
\label{fig1}
\end{figure*}

\begin{figure*}[htbp]
\centerline{\includegraphics[width=0.8\textwidth]{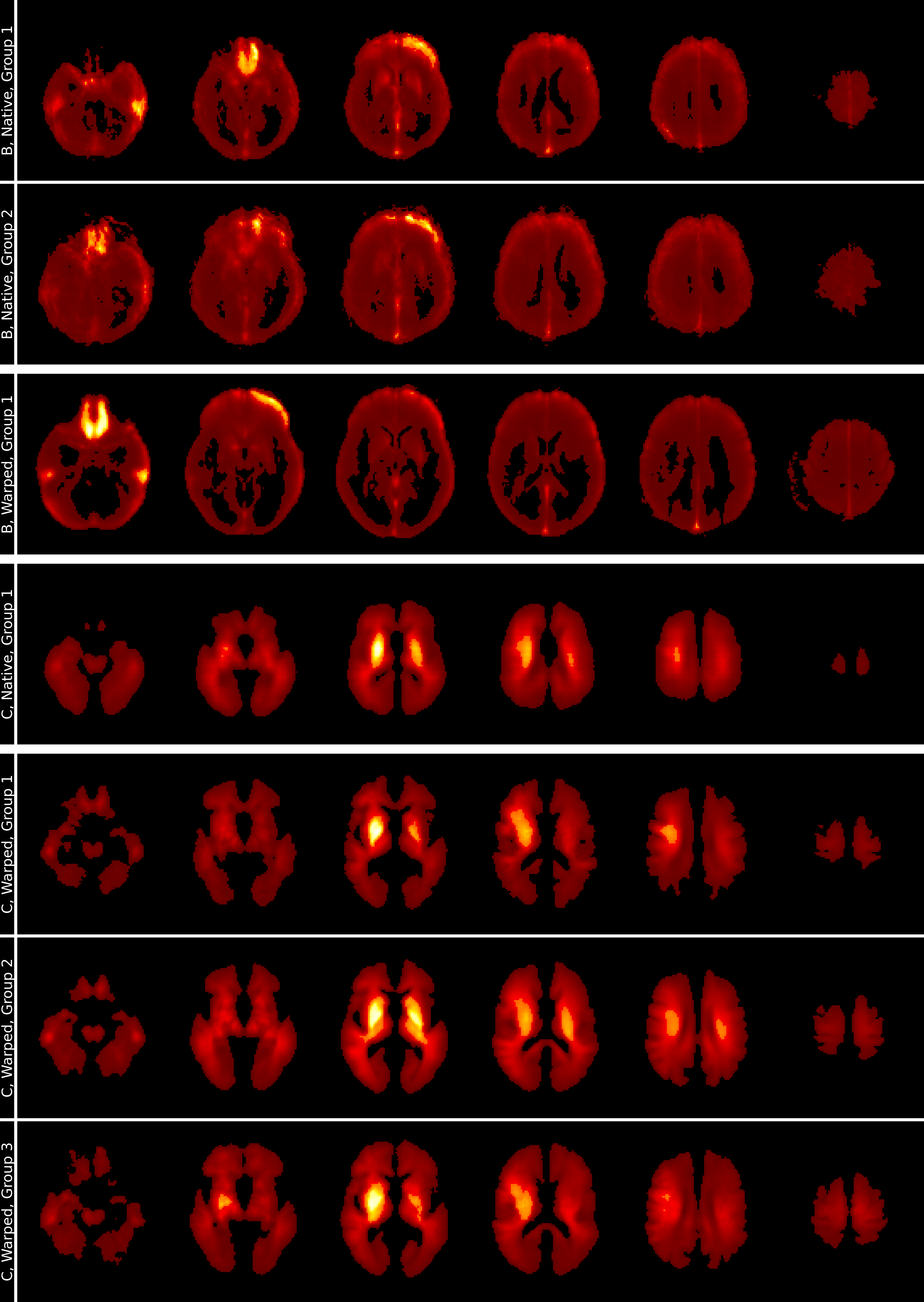}}
\caption{Mean heatmaps for the groups identified using spectral clustering and eigenvalue analysis for models B and C. Model C shows a clear separation for warped heatmaps between Group 1 (row 5, all heatmaps are from AD) and Group 2 (row 6, most heatmaps are from NC). Model C highlights the left and the right basal ganglia, suggesting a more structured relevance pattern compared to Model B. Images are shown in standard-radiological view, causing the left and right side of the brain to be flipped.}
\label{fig2}
\end{figure*}

We visualized the clustering of the heatmaps in native subject space and in MNI152 space using t-SNE, initialized with the normalized, symmetric, and positive semi-definite Laplacian matrix derived from the spectral clustering affinity matrix. Fig.~\ref{fig1} illustrates the grouping of heatmaps and corresponding warped heatmaps for all models. Group mean heatmaps, based on spectral clustering groupings, are presented in Fig.~\ref{fig2}.

\section{Discussion}

Our study extends previous research on the interpretability of deep learning models for AD classification using R2* maps. Earlier analyses \cite{tinauer_explainable_2024} showed that CNNs primarily focus on relaxation rate changes in the basal ganglia. Here, we advance this understanding by applying spectral clustering to LRP-derived heatmaps, enabling a more structured assessment of decision patterns and uncovering systematic classification strategies beyond single-instance explanations.

The t-SNE visualization in Fig.~\ref{fig1} shows that heatmap warping before spectral clustering influences the grouping (row 1 vs. row 2). Notably, only model C, which applies relevance-guided training and warping, exhibited heatmap clustering aligning with subject groups (NC vs. AD). This suggests that models trained without explicit spatial constraints (models A and B) may rely on less structured features, while relevance-guided training helps capture more biologically meaningful patterns.

Mean heatmaps in Fig.~\ref{fig2}, derived from spectral clustering groupings, reveal consistent relevance patterns that offer insight into the classifier's decision strategies for models B and C. In the best-performing model (model C), which applies the relevance-guided approach, the separation between AD and NC predictions is more pronounced compared to models A and B. The mean warped heatmap for Group 1 (all AD) in model C shows greater relevance in the right basal ganglia, while the Group 2 (mostly NC) heatmap attributes equal relevance to both the left and right basal ganglia. This suggests that the relevance-guided approach enhances the model’s ability to focus on meaningful features, providing a more interpretable decision strategy. Interestingly, Group 3 also highlights the left basal ganglia and surrounding tissue, but the group contains heatmaps from both NC and AD subjects, potentially reflecting structural differences and nonlinear registration. This shows the need for further exploration of these learned representations. In contrast, model B's heatmaps in native space show more reliance on brain masks and volume influences, with varying highlighted regions. After warping to the MNI152 space, these groups merge, emphasizing that postprocessing may obscure position-driven features.

The absence of significant performance differences between models B and C indicates that classification accuracy alone is insufficient to assess model robustness. Despite comparable performances across models, spectral clustering revealed distinct decision patterns, highlighting the importance of explainability techniques for evaluating model reliability. Importantly, our findings emphasize that seemingly minor training choices -such as skull-stripping or relevance-guided regularization- can substantially impact learned representations.

\section{Conclusion}

This study utilized quantitative MRI data (R2*) for deep learning classification and layer-wise relevance propagation (LRP) in a clinical cohort of Alzheimer’s disease patients. By extending our previous research on heatmapping validation \cite{tinauer_explainable_2024} by integrating chemical \cite{langkammer_quantitative_2010} and in-vivo \cite{damulina_cross-sectional_2020} iron mapping studies, our findings confirm that heatmapping approaches can serve as valuable tools to identify areas of tissue changes and provide deeper insights into the internal mechanisms of deep learning-based classification networks. Spectral clustering applied to LRP-based heatmaps allowed us to evaluate classifier decision strategies systematically. Future studies are needed to further explore the influence of preprocessing artifacts on model decisions.

\bibliographystyle{IEEEtran}
\bibliography{main.bib}

\end{document}